# Thermal conductivity spectrum calculation from first-principles-based harmonic phonon theory

Takuma Shiga[1], Daisuke Aketo[1], Lei Feng[1], and Junichiro Shiomi[1,2,a]

[1]*Department of Mechanical Engineering, The University of Tokyo, 7-3-1 Hongo, Bunkyo, Tokyo, 113-8656, Japan*
[2]*Center for Materials research by Information Integration, National Institute for Materials Science, 1-2-1 Sengen, Tsukuba, Ibaraki, 305-0047, Japan*

a) Email address: shiomi@photon.t.u-tokyo.ac.jp

In recent years, nanostructuring of dielectric and semiconducting crystals has enhanced controllability of their thermal conductivity. To carry out computational material search for nanostructured materials with desirable thermal conductivity, a key property is the thermal conductivity spectrum of the original single crystal, which determines the appropriate length scale of nanostructures and mutual adaptability of different kinds of nanostructures. Although the first-principles phonon transport calculations have become accessible, the anharmonic lattice dynamics calculations are still heavy to scan many materials. To this end, we have developed an empirical model that describes the thermal conductivity spectrum in terms only of harmonic phonon properties and bulk thermal conductivity. The model was tested for several crystals with different structures and thermal conductivities, and was confirmed to reproduce the overall profiles of thermal conductivity spectra and their accumulation functions obtained by the first-principles anharmonic calculations.

Recent progress in nanostructuring has realized large controllability in thermal conductivity of crystalline dielectrics and semiconductors. In the scope of realizing low thermal-conductivity materials, by forming nanowires, superlattices, or nanocrystallines with inter-surface/interface distance smaller than the effective phonon mean free path (MFP), surface/interface phonon scattering has been shown to reduce thermal conductivity even to values close to that of amorphous allotropes ("amorphous limit")[1-3]. This has been applied widely to materials development, especially for thermoelectrics, whose figure-of-merit increases by lowering the thermal conductivity[4-10]. Here, a key challenge to further improve or design the materials is to predict the relation between the thermal conductivity reduction and the representative length scales (inter-surface/interface distances) of the nanostructures, however, that requires the knowledge in mode-dependent contribution to thermal conductivity, i.e. thermal conductivity spectrum.

Here, it is important to note that phonons with a wide range of frequencies (energies) or MFPs can contribute to thermal conductivity, unlike electrical conductivity that can be attributed to transport of electrons and holes around Fermi surface. For instance, in single-crystal silicon, transports of phonons with frequency up to 16 THz and MFPs up to 10 μm noticeably contribute to heat conduction[11]. The fact that MFPs can vary by orders of magnitude makes the heat conduction in nanostructured materials greatly multi-scale: phonons with MFP smaller (diffusive phonons), similar (quasi-diffusive phonons), and larger (ballistic phonons) than the representative length scale of nanostructures are all present, and are affected differently by the surface/interfaces. This also means combining multiple kinds of nanostructures with different representative lengths and corresponding scattering frequency ranges can reduce thermal conductivity more than a single kind, and such mutual adoptability of nanostructures with different spectral characteristics has been numerically demonstrated[12-14].

This necessity to consider heat conduction in terms of spectrum has led to a great advance in techniques to quantify the thermal conductivity spectrum. On the experimental side, measurement methods such as time-domain thermoreflectance (TDTR) and broad-band frequency-domain thermoreflectance (BB-FDTR) methods have realized thermal transport measurements within the region below some phonon MFPs of crystals, and, through the variation of the region size, have been used to obtain the thermal conductivity spectrum[15-21]. As for the calculation techniques, development of first-principles methods to calculate the anharmonic interatomic force constants (a-IFCs) has realized accurate anharmonic lattice dynamics calculations of mode-dependent phonon relaxation times (First-principles-based anharmonic lattice dynamics, FP-ALD), which, together with harmonic dispersion relations, give phonon MFPs and the thermal conductivity spectrum from Boltzmann transport viewpoint[11,22,23]. Similar properties can be also obtained from molecular dynamics simulations using the same a-IFCs[23]. Furthermore, thermal conductivity reduction by nanostructures can be predicted by inputting the mode-dependent phonon transport properties into simple scattering models[12,24] and/or phonon Monte Carlo simulations[13,25].

The objective of this letter is in line with an effort to use such knowledge in thermal conductivity spectrum to search for nanostructured materials with low thermal conductivity. Whether to base the search on physical mechanism or machine learning (materials informatics)[26-28], it would require a large data set, i.e. phonon transport properties and thermal conductivity spectra of many materials. Although, in theory, this could be done using the above FP-ALD, the cost for first-principles calculations of a-IFCs is still high, particularly for materials with complex structures[29-31]. Note that although anharmonic lattice dynamics calculation has been successfully applied to many materials[11,22,23,29-35] and has become accessible through the open codes[36-38], there are details (often not described in the papers) such as dependence on the cutoff distances of the a-IFCs, which needs to be carefully examined for each material. The care is particularly important for mode-dependent anharmonic relaxation times because validation with experiments such as inelastic neutron scattering is yet hardly quantitative[39].

To realize faster calculations of accurate thermal conductivity spectrum, in this work, we have developed a method to empirically calculate thermal conductivity spectrum of a crystal by only using harmonic phonon transport properties and the value of bulk (total) thermal conductivity. We have previously reported an empirical scaling of thermal conductivity spectrum using harmonic phonon property and bulk thermal conductivity[40], which is useful to identify the shortest and longest MFPs with noticeable contribution to thermal conductivity, but this work further challenges to obtain the overall profile of the thermal conductivity spectrum. The first-principle calculation of the harmonic properties by lattice dynamics can be done much faster than that of anharmonic properties, and bulk thermal conductivity values are known or can be measured by experiments. The calculation of harmonic properties is much less burdening than that of anharmonic ones also because their validity can be verified by comparison with the experimentally measured dispersion relations. Although some models have been previously proposed to empirically model lattice thermal conductivity such as Callaway-Holland model[41,42], since they only describe transport of long-wavelength acoustic phonons, they cannot be used to obtain the overall thermal conductivity spectrum.

We express lattice thermal conductivity based on kinetic gas model[43]:

$$\kappa = \frac{1}{3\Omega} \sum_{k,j} C_j(k) v_j^2(k) \tau_j(k), \quad (1)$$

where $C_j(k)$, $v_j(k)$, and $\tau_j(k)$ are modal heat capacity, group velocity, and relaxation time of phonon with wavevector $k$ and phonon branch $j$. $\Omega$ is the volume of the system, and the factor "3" comes from isotropy. By converting the $k$-space summation to frequency($\omega$)- or MFP($\Lambda$)-space integration, we obtain,

$$\kappa = \frac{1}{3\Omega} \sum_j \int_0^\infty D_{j,\omega} C_{j,\omega} v_{j,\omega}^2 \tau_{j,\omega} \, d\omega$$
$$= \frac{1}{3\Omega} \sum_j \int_0^\infty D_{j,\Lambda} C_{j,\Lambda} v_{j,\Lambda} \Lambda_j d\Lambda. \quad (2)$$

Note that the conversion of integral space yields density of states $D_{\omega/\Lambda}$[44].

In this work, we focus only on the three-phonon scattering that is the main source of intrinsic thermal resistance in single crystals. By applying the first-order perturbation theory to third-order lattice anharmonicity, phonon relaxation time, $\tau$, is given as[11,23,43]

$$\tau_j^{-1}(k) = \frac{2\pi}{N_k} \sum_{1,2} \frac{|V_3(kj,k_1j_1,k_2j_2)|^2}{\omega \omega_1 \omega_2}$$
$$\times \begin{bmatrix} (n_1 - n_2)\delta(\omega + \omega_1 - \omega_2) + \\ \frac{1}{2}(n_1 + n_2 + 1)\delta(\omega - \omega_1 - \omega_2) \end{bmatrix}, \quad (3)$$

where $n$ and $N_k$ are the Bose-Einstein distribution and number of $k$-points, respectively. Subscripts 1 and 2 represent phonon modes; $(k_1,j_1)$ and $(k_2,j_2)$. $V_3$, measure of anharmonic scattering magnitude, is defined as

$$V_3(kj,k_1j_1,k_2j_2) =$$
$$\frac{1}{N_k} \left(\frac{\hbar}{2}\right)^{3/2} \sum_{\{l,\eta,\alpha\}} \frac{e_j^\alpha(k|\eta) e_{j_1}^\beta(k_1|\eta_1) e_{j_2}^\gamma(k_2|\eta_2)}{\sqrt{M_\eta M_{\eta_1} M_{\eta_2}}} \quad (4)$$
$$\times \Psi_{l\eta,l_1\eta_1,l_2\eta_2}^{\alpha\beta\gamma} e^{i(k \cdot l + k_1 \cdot l_1 + k_2 \cdot l_2)},$$

where $M$, $e$, and $\Psi$ denote atomic mass, eigenvector, and third-order a-IFCs. The summation with respect to primitive unit vector ($l$), atomic index in primitive unit cell ($\eta$), and Cartesian component ($\alpha$) is non-zero as long as $k+k_1+k_2$ is commensurate with reciprocal vector $G$, which corresponds to the momentum conservation.

Firstly we will consider the phonon populations, $(n_2-n_1)$ and $(n_1+n_2+1)$. At the high temperature limit, the Bose-Einstein distribution $n$ becomes $k_BT/\hbar\omega$, where $T$ is the temperature. Note that this classical approximation is reasonable for most of materials with moderate thermal conductivity (e.g. thermoelectric materials) above room temperature. By considering also the energy conservations in the Dirac delta functions ($\omega=\omega_1\pm\omega_2$), the phonon population terms are simply proportional to $(\omega/\omega_1\omega_2)T$, which leads to

$$\tau_j^{-1}(k) = \frac{2\pi k_B T}{\hbar N_k} \sum_{1,2} \frac{|V_3(kj,k_1j_1,k_2j_2)|^2}{\omega_1^2 \omega_2^2}$$
$$\times \left[ P_3^{(+)}(kj|k_1j_1,k_2j_2) + \frac{1}{2}P_3^{(-)}(kj|k_1j_1,k_2j_2) \right]. \quad (5)$$

where $P_3$ is the scattering phase space proposed by Lindsay and Broido[45];

$$P_3^{(\pm)}(kj|k_1j_1,k_2j_2) = \delta(\omega \pm \omega_1 - \omega_2)\delta_{k\pm k_1,k_2+G}. \quad (6)$$

This is basically the two-phonon density of states accounting not only for energy (frequency) conservation but also for momentum conservation. Therefore, it characterizes the number of phonon scattering channels.

It can be understood from Eq. (5) that the relaxation time basically is a product of the strength (scattering magnitude) and number ($P_3$) of the scattering processes, which are anharmonic and harmonic properties, respectively. The current modeling is based on the assumption that the profile (shape) of thermal conductivity (or relaxation time) spectrum is determined mainly by the scattering phase space. Therefore, while rigorously computing the scattering phase space by first principles, we adopt a simplification for the scattering magnitude. For this, we use the Klemens model ($V_3 \propto \omega\omega_1\omega_2$)[46,47], which has been extensively employed to model the intrinsic phonon-phonon scattering. This leads to the following expression for the relaxation time,

$$\tau_j^{-1}(k) = A\omega_j^2(k)T$$
$$\times \sum_{1,2} \left[ P_3^{(+)}(kj|k_1j_1,k_2j_2) + \frac{1}{2}P_3^{(-)}(kj|k_1j_1,k_2j_2) \right]. \quad (7)$$

Here the unknown parameter $A$ characterizing the scattering magnitude is left as a tuning parameter, which can be obtained by matching $\kappa$ in Eq. (1) to the known value of bulk thermal conductivity ($\kappa_{\text{bulk}}$).

Figure 1 shows frequency-dependent phonon relaxation times of several materials at 300 K calculated from Eq. (7). To check the applicability of the model to a wide range of materials, we have chosen materials with different structural complexities (number of atoms in the primitive cell ranging from 2 to 17) and thermal conductivities (ranging from 1 to 100 W/m-K). The figure also shows the results of FP-ALD for comparison. It is noted that the data for unfilled- and filled-skutterudites ($CoSb_3$ and $BaCo_4Sb_{12}$) were taken from Ref. 29. In the scattering phase space calculation, the integration over the first Brillouin zone in the Eq. (7) was done by using the tetrahedron method. Harmonic and anharmonic IFCs were calculated using density functional theory (DFT) and *real-space displacement* method[11,48]. The DFT calculation conditions have been described elsewhere[11,23,29,30,33,35,40], and all calculations were done by using the ALAMODE package[36]. In the calculation of scattering phase space, we used $N_k \times N_k \times N_k$ uniform mesh and chose $N_k$ depending on the material ($N_k$=30 for Si, PbTe, $Mg_2Si$, and ZrCoSb, $N_k$=20 for $SrTiO_3$, $N_k$=10 for $CoSb_3$ and $BaCo_4Sb_{12}$). It has been confirmed that calculating Eq. (3) with quantum and classical phonon distributions results only in minute difference in the profiles of thermal conductivity spectra at 300 K when normalized with $\kappa_{bulk}$.

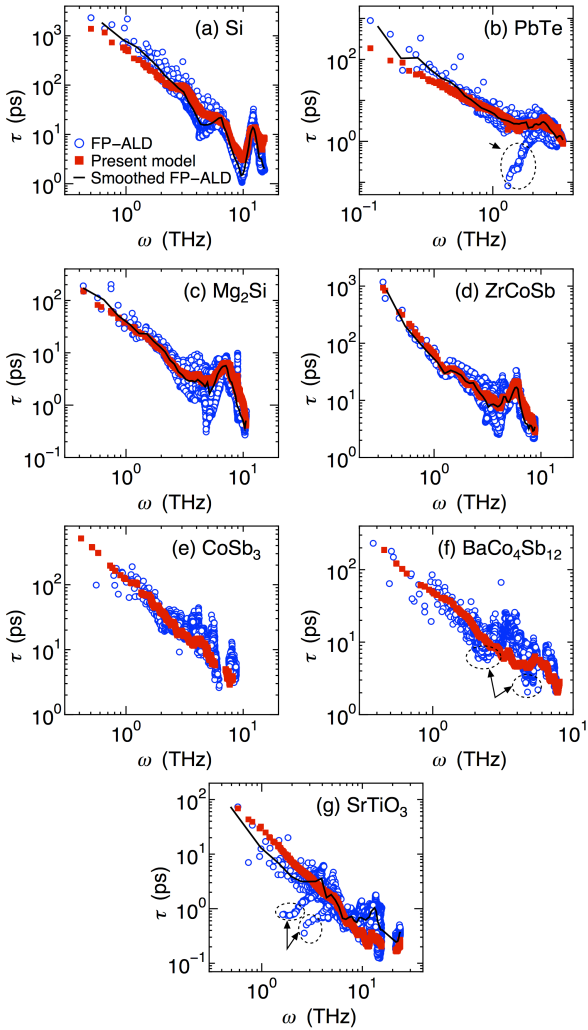

**Fig. 1** Frequency-dependent phonon relaxation time, $\tau(\omega)$, at 300 K of (a) Si, (b) PbTe, (c) $Mg_2Si$, (d) ZrCoSb, (e) $CoSb_3$, (f) $BaCo_4Sb_{12}$, and (g) $SrTiO_3$. Blue open circles and red filled squares are results of FP-ALD and model proposed in this work, respectively. Solid black lines denote FP-ALD results smoothed in frequency regime. FP-ALD results for $CoSb_3$ and $BaCo_4Sb_{12}$ are taken from Ref. 29.

For all the materials in Fig. 1, the frequency-dependent relaxation times obtained by the current model fall within the profiles obtained by FP-ALD. The fluctuation of the profile is much larger in the FP-ALD results, especially in the intermediate frequency regime. This is due to the branch dependence of the scattering magnitude that is not incorporated in the current model. However, a good agreement can be confirmed when comparing the model with FP-ALD results smoothed in frequency domain (solid black line in Fig. 1), where the frequency window of smoothing is set to 1/50 of the maximum frequency. The smoothed profiles are not shown for $CoSb_3$ and $BaCo_4Sb_{12}$ because the assignment of the relaxation time to each mode is not available in Ref. 29, however, it can be seen that data obtained by the current model lie in the middle of the ones by FP-ALD. It is interesting that the detail features of the profile such as the complex oscillation in the high frequency regime are successfully captured by the model. Since FP-ALD relaxation times with and without the smoothing give nearly identical thermal conductivity spectrum, we conclude from the agreement that the current model well reproduces the frequency-dependent relaxation times.

On the other hand, the current model does fail to capture some detail features. One example is the anomalously small relaxation times of some transverse optical modes in PbTe and $SrTiO_3$ denoted with dotted circles in Fig. 1(b) and Fig. 1(g). PbTe is known to have nearly ferroelectric transverse optical modes at around zone center with extremely large anharmonicity (diverging mode Grüneisen parameter)[33,49,50], which has been also observed in inelastic neutron scattering experiments[51,52]. $SrTiO_3$, in addition to the similar ferroelectric modes, has an anti-ferroelectric mode around the R-point, which gives rise to large anharmonicity[35]. Another example is the small relaxation times at 2.4 THz and 4.8 THz in $BaCo_4Sb_{12}$, which are highlighted with dotted circles in Fig. 1(f). In $BaCo_4Sb_{12}$, introduction of Ba atom into a free space in $CoSb_3$ host lattice changes the lattice anharmonicity and scattering phase space of the lattice surrounding the Ba atom, which leads to reduction of the relaxation times (*rattling effect*)[29,30]. This local anharmonic feature cannot be reproduced by the model because it does not account for the polarization-dependent variation in anharmonicity. However, the reduction in the relaxation times is limited to narrow frequency regimes and the influence on the overall profile is minute. Therefore, by incorporating the effect of rattling modes to the overall reduction of relaxation time (by three-phonon scattering with other modes) through the fitting parameter $A$ in Eq. (7), the model successfully reproduces the thermal conductivity spectrum of $BaCo_4Sb_{12}$ as it will be shown later (Fig. 2 and 3).

One last feature worth being discussed is the power law dependence of the relaxation time in low frequency regime, which has been extensively discussed in the literature[11,40,44,53]. From the slopes of the profiles in Fig. 1, we can see that the current model in general successfully reproduces of the power law dependence but the residual is smaller for certain materials than others. The ones with relatively large residual are Si and PbTe, where the model somewhat underestimates the FP-ALD relaxation times in the low frequency regime. While the power law $\omega^{-2.1}$ ($\omega^{-2.4}$) is found by fitting the FP-ALD data of Si (PbTe) in the low frequency regime, the same fitting for the data obtained by the current model gives $\omega^{-1.8}$ ($\omega^{-1.7}$). Note that the estimated power law deviates from $\omega^{-2.0}$ in Eq. (7) due to the frequency dependences of scattering phase space. Exact reproduction of the power-law exponent would require incorporation of branch-dependent scattering magnitude that are neglected in the Klemens model, which remains as a future task.

We next evaluate thermal conductivity spectrum. The integrands in Eq. (2) are thermal conductivity spectra in frequency ($\kappa_\omega$) and MFP spaces ($\kappa_\Lambda$). Here, heat capacity and phonon group velocities are calculated from the aforementioned lattice dynamics calculations using the harmonic IFCs obtained from first principles. Fig. 2 shows the obtained $\kappa_\omega$ in comparison with those from FP-ALD calculations. As seen in the figure, the current model reproduces $\kappa_\omega$ of the FP-ALD calculations, with particularly good agreement in intermediate to high frequency regimes. For more conservative comparison (less attention to reproduction of the peaks), we have also calculated the cumulative thermal conductivities (CTC), accumulation functions of the thermal conductivity spectra (See Supplementary Materials[54] for the expressions of CTC). As shown in Fig. 3, the overall profiles of CTC with respect to frequency and MFP agree well with those of FP-ALD.

Although the current model does not reproduce the local fluctuations in relaxation times due to polarization-dependent scattering magnitude (Fig. 1), this has minor effect on the reproduction of thermal conductivity spectrum because the profile is smoothed by being multiplied with heat capacities and squared group velocities. Some discrepancies in low-frequency and long-MFP regimes, which are noticeable for Si and PbTe, are due to the above mentioned underestimation of the power-law exponent in the frequency dependence of relaxation time. It is worth mentioning that correction of power-law exponent in Eq. (7) can improve the CTC profiles. For example, adopting the power-law exponent $\omega^{-2.5}$ instead of $\omega^{-2.0}$ recovers accurate CTC profiles in low-frequency and long-MFP regimes as shown by black dashed lines in Figs. 2(a)-(b) and 3(a)-(b) (See Fig. S3 in the Supplementary Materials[54] for the modified relaxation time profiles).

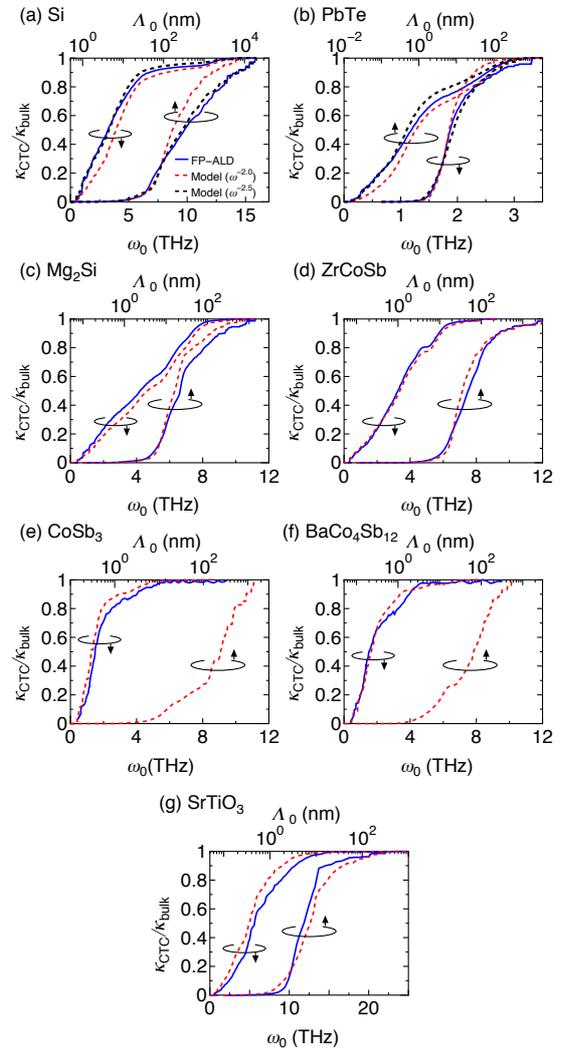

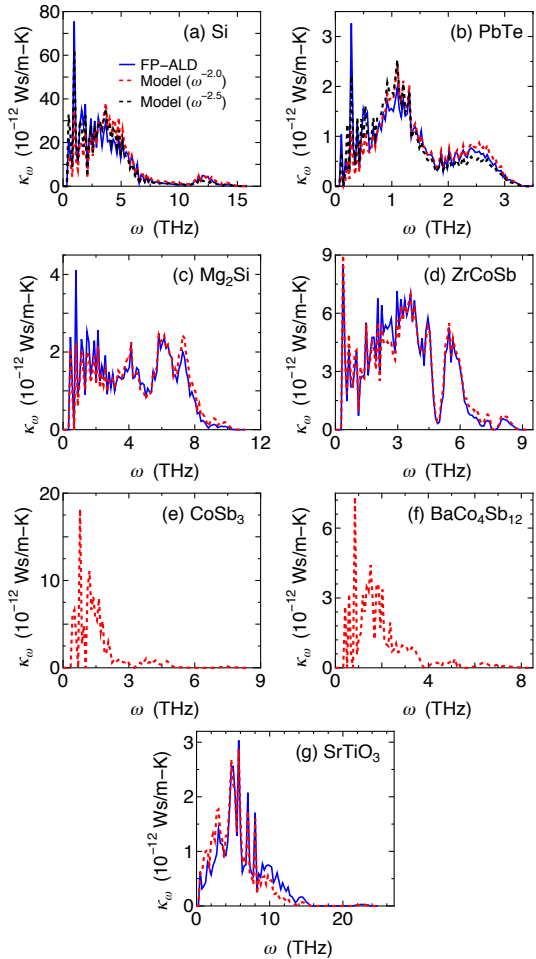

**Fig. 2** Thermal conductivity spectra in frequency space ($\kappa_\omega$) of (a) Si, (b) PbTe, (c) $Mg_2Si$, (d) ZrCoSb, (e) $CoSb_3$, (f) $BaCo_4Sb_{12}$ and (g) $SrTiO_3$ at 300 K. Blue solid and red dashed lines are results of FP-ALD and present model.

**Fig. 3** Cumulative thermal conductivities (CTC) in frequency and MFP spaces of (a) Si, (b) PbTe, (c) $Mg_2Si$, (d) ZrCoSb, (e) $CoSb_3$, (f) $BaCo_4Sb_{12}$ and (g) $SrTiO_3$ at 300 K. Blue solid and red dashed lines are results of FP-ALD and the current model. FP-ALD values for $CoSb_3$ and $BaCo_4Sb_{12}$ are taken from Ref. 29. Black dashed lines are model calculations with $\omega^{-2.5}$ instead of $\omega^{-2.0}$ in Eq. (7). The expression of CTC is described in Eq. (S1) in the Supplementary Materials[54].

In summary, we have developed an empirical model that describes thermal conductivity spectrum of a crystal in terms only of harmonic phonon properties and the bulk thermal conductivity. In the model, while harmonic phonon properties (heat capacity and group velocity) are accurately calculated by first-principles-based (harmonic) lattice dynamics, relaxation time is modeled by the scattering phase space (momentum-conserving two-phonon density of states) and the Klemens model at the classical limit. By validating the model against the first principles anharmonic lattice dynamics calculations for several materials with different structural complexities and thermal conductivities, we have shown that the model successfully reproduces the overall profiles of thermal conductivity spectra and their accumulation functions in both frequency and MFP domains. Although better estimation of the exponent of power-law frequency dependence of relaxation time and generalization to anisotropic materials remains as future task, the facileness of the calculation only requiring the harmonic phonon properties and bulk thermal conductivity enables us to calculate thermal conductivity spectra of many materials, leading to computational material search for nanostructured materials with desirable thermal conductivity.

**Acknowledgements**

This work was partially supported by KAKENHI (Grand Numbers 15K17982, 26709009, and 26630061), the Thermal and Electrical Energy Technology Foundation, and MI$^2$I: "Materials research by Information Integration" Initiative program supported by Japan Science and Technology Agency. This work was performed using facilities of the Institute for Solid State Physics, the University of Tokyo, and the TSUBAME2.0 supercomputer in the Tokyo Institute of Technology supported by the MEXT Open Advanced Research Facilities Initiative.